\documentclass[9pt,twocolumn,twoside]{opticajnl}
\journal{opticajournal} 

\setboolean{shortarticle}{true}

\usepackage{color,soul}
\usepackage{amsmath}
\usepackage{lineno}

\title{Next-generation soliton frequency combs in photonic-crystal and nanocomposite microresonators}

\author[1,2,*]{Haixin Liu}
\author[1]{Alexa Carollo}
\author[1,2]{Jizhao Zang}
\author[1,2]{Scott B. Papp}

\affil[1]{Time and Frequency Division, National Institute of Standards and Technology, 325 Broadway, Boulder, Colorado 80305, USA}
\affil[2]{Department of Physics, University of Colorado, Boulder, Colorado 80309, USA}

\affil[*]{Haixin.Liu@colorado.edu}

\begin{abstract}
Microresonator frequency combs offer tremendous opportunity to advance applications in fundamental research and technology by linking the optical and microwave frequency domains. Kerr-nonlinear microresonators further enable the generation of portable, integrated optical frequency combs, which are called microcombs. However, the dispersion engineering usually suffers from the small, geometric parameter space, and achieving soliton microcombs is challenging and usually requires complicated experimental techniques and setups. In recent years, the invention of photonic-crystal resonators (PhCRs) provides access to solitons in a convenient and stable way while its mechanism has not been fully understood. In this article, we highlight the perspectives of generating solitons for various applications in PhCRs and give a thorough understanding of the dynamics of soliton formation. We also propose a nanocomposite waveguide structure for the optimization of group velocity dispersion (GVD), lifting the limitation on geometric parameter space. We  apply it to the design of the pump-harmonic microcomb, a new concept for f-2f self-referencing in microcombs. Inspired by the pulse-driven microresonator, we provide a scheme of two-microresonator network, retaining the convenience of using a continuous wave laser for microcomb generation. Our work depicts a blueprint of achieving the next generation soliton frequency combs in PhCR and nanocomposite microresonators and highlights their great prospects in optical metrology, precision measurement and optical data transmission.
\end{abstract}

\setboolean{displaycopyright}{false} 

\begin{document}

\maketitle

\section{Introduction}
Optical frequency combs provide a direct link between the optical and microwave domains, enabling a powerful tool for precision metrology \cite{RevModPhys.78.1279}. Since their inception, optical frequency combs have enabled groundbreaking developments in fundamental research including optical atomic clocks \cite{newman2019architecture}, high-resolution spectroscopy \cite{suh2016microresonator,picque2019frequency}, astronomy research \cite{obrzud2019microphotonic,metcalf2019stellar} and quantum state generation \cite{reimer2016generation}, as well as engineering applications like optical data transmission \cite{Pirmoradi_2025,Daudlin_2025,fulop2018high,yang2022multi,deakin20242} and precision distance measurement \cite{coddington2009rapid,trocha2018ultrafast}. Traditional frequency combs are generated by mode-locked femtosecond lasers and have bulky and power-consuming setups. This has driven significant interest in miniaturized, scalable alternatives that can have similar performance with portability. In the past twenty years, microresonator-based frequency combs, or microcombs, have emerged as one of the most promising solutions to this challenge \cite{kippenberg2011microresonator}. Relying on the Kerr nonlinearity, solitons with frequency comb spectra can be excited through four-wave mixing interaction (FWM) in high-Q optical microresonators pumped by a continuous wave laser. As the light is highly concentrated in the wavelength-scale waveguides, the nonlinear interaction is greatly enhanced, resulting in a much lower operating threshold ($\sim$ mW) \cite{lu2019milliwatt}. A major advantage of microcombs is their amenability to wafer-scale fabrication and future integration with complementary metal-oxide-semiconductor (CMOS) circuits for hybrid photonic-electronic systems \cite{xie2022soliton}. Moreover, the promising application of microcombs in photonic AI computing accelerators has been demonstrated in recent years \cite{feldmann2021parallel}. So far, microcombs have been demonstrated in platforms of various materials including silicon nitride \cite{jahanbozorgi2023generation}, tantala \cite{spektor2024photonic}, aluminum gallium arsenide \cite{shu2022microcomb}, chalcogenide \cite{xia2023integrated}, and lithium niobate \cite{wang2019monolithic}.

A critical factor that determines the microcomb spectrum is the integrated group velocity dispersion (GVD) of the microresonator, which is defined by $D_{\text{int}} = \nu_{\mu} - (\nu_0+\text{FSR}\mu)$, where $\mu$ denotes the mode number relative to the pump and $\nu_{\mu}$ is its resonance frequency of the cold cavity, and FSR represents the free spectral range at the pump frequency \cite{fujii2020dispersion}. By designing the geometrical parameters including the ring width (RW) and thickness ($t$) of microresonators, we can tailor $D_{\text{int}}$ and thus control the spectra of the microcomb. This process is called dispersion engineering \cite{Yang_2016}. In an anomalous GVD microresonator with higher-order dispersion, coherent emission of dispersive waves (DWs) can arise around the modes with vanishing $D_{\text{int}}$ \cite{brasch2016photonic}. For an octave-spanning microcomb that has a long wavelength DW (LWDW) and a short wavelength DW (SWDW) spanning an octave, its carrier-envelope frequency $f_\text{ceo}$ can be detected and stabilized through f-2f self-referencing, which further makes a microcomb into a robust phase-stabilized frequency standard, greatly promoting its usages \cite{brasch2017self}. Recently, foundry manufacturing of octave spanning microcombs at volume has been demonstrated, highlighting its prospects in widely accessible applications \cite{zang2024foundry}. 

While dispersion engineering enables the control of the microcomb spectrum, the generation of the soliton is always a challenge. For an anomalous GVD microresonator, a chaotic regime will precede the existence regime of the steady-state soliton, preventing its generation in experiments. This issue was traditionally tackled by various techniques, such as using an auxiliary laser for thermal compensation during the chaos-soliton transition, using a single-sideband suppressed-carrier frequency shifter to fast sweep the pump frequency or implementing self-injection locking, but at the cost of a much more complicated experimental setup \cite{zhang2019sub,li2017stably,PhysRevA.106.053508}. In other cases, a pulsed laser instead of a continuous-wave laser is used as the pump to excite the soliton \cite{Anderson_2021}. On the other hand, in a normal GVD microresonator, the Kerr shift cannot compensate the $D_{\text{int}}$ profile, disabling the phase matching and thus the generation of the primary comb. This results in a considerable limitation in the available geometric parameter space of microresonators for soliton generation.

Recently, photonic-crystal resonators (PhCRs) have emerged as powerful tools for the mode-specific manipulation of dispersion. By implementing a periodic sidewall modulation on the waveguide, we can split a chosen resonance mode into two standing-wave modes, and the split can be precisely controlled by adjusting the amplitude of the sidewall modulation $A_{\text{PhC}}$ without any change in the waveguide dimensions. This technique provides extra flexibility in phase matching  \cite{black2022optical}. In the past few years, optical parametric oscillators with output over $40$ mW and a tunable span over an octave have been demonstrated in PhCRs \cite{liu2024threshold,Stone_2023,brodnik2025activatinghighpowerparametricoscillation,brodnik2025nanopatterned}. Recently, high quality factor (up to $1.2\times10^7$) PhCRs manufactured by a commercial foundry on 300 mm silicon wafers were reported, suggesting their perspectives in much broader applications \cite{liu2025implementing}. More importantly, sufficient experimental evidence shows that PhCRs enable the stable generation of broadband solitons in anomalous GVD and high pump-to-comb efficiency solitons in normal GVD without extra techniques, greatly reducing the experimental complexity and the limitation in dispersion engineering \cite{yu2021spontaneous,Zang_2025,Jin_2025}. So far, however, a comprehensive understanding of why solitons form more easily in PhCRs has not been specifically explored. Therefore, a full understanding of the dynamics underlying the soliton formation in PhCR is essential to the future design of the next-generation microcombs.

In addition to PhCRs, we propose the concept of the pump-harmonic microcomb as an improved alternative to the existing few examples of octave-spanning microcombs. This new pump harmonic case leverages the pump wave and the SWDW with an octave span for self-referencing, avoiding the issue that the power of LWDW is not high enough for frequency doubling. This new kind of self-referencing microcomb requires much more favorable GVD that supports sufficient power of the SWDW. To address the challenge in dispersion engineering, we propose a revolutionary scheme by use of a nanocomposite microresonator which has a multilayer structure made of tantala (Ta$_2$O$_5$) and silica (SiO$_2$). The additional degrees of freedom, such as the thickness of individual layers, enable the simultaneous optimization of the second-order dispersion $D_2$ and the frequency of the LWDW ($\nu_\text{L}$), which cannot be achieved with the conventional waveguide structure. These nanocomposite devices can be realized through ion-beam sputtering and typical nanofabrication.

In this article, we discuss the prospects of soliton frequency combs in photonic-crystal and nanocomposite microresonators in future photonic systems. We provide an overview of PhCRs including their potential applications in integrated photonic circuits and their characterization in experiments. The mechanism of soliton formation is explained through theoretical analysis and numerical simulation for both anomalous and normal GVD PhCRs, and stable soliton generation is demonstrated in experiments. We also explore the scheme employing a nanocomposite microresonator formed by Ta$_2$O$_5$ and SiO$_2$ to generate pump-harmonic microcombs. Our simulation shows that this nanocomposite architecture provides a revolutionary improvement in the dispersion profile that cannot be achieved by any conventional waveguide and results in a 20 dB increase in the SWDW power. The idea also applies to the dispersion engineering of general waveguides. We further propose a two-stage microresonator network. With a PhCR and the second nanocomposite microresonator cascading, this network can achieve stable pump-harmonic microcomb generation with improved SWDW power. Our work highlights the prospects and new directions for next-generation microcombs and contributes to the broader effort of building compact, robust, and scalable frequency standards for future integrated photonic systems.

\section{AN OVERVIEW OF PHCRS}
\subsection{PhCRs in Integrated Photonics Circuits}

\begin{figure}[ht]
\centering
\includegraphics[width=\linewidth]{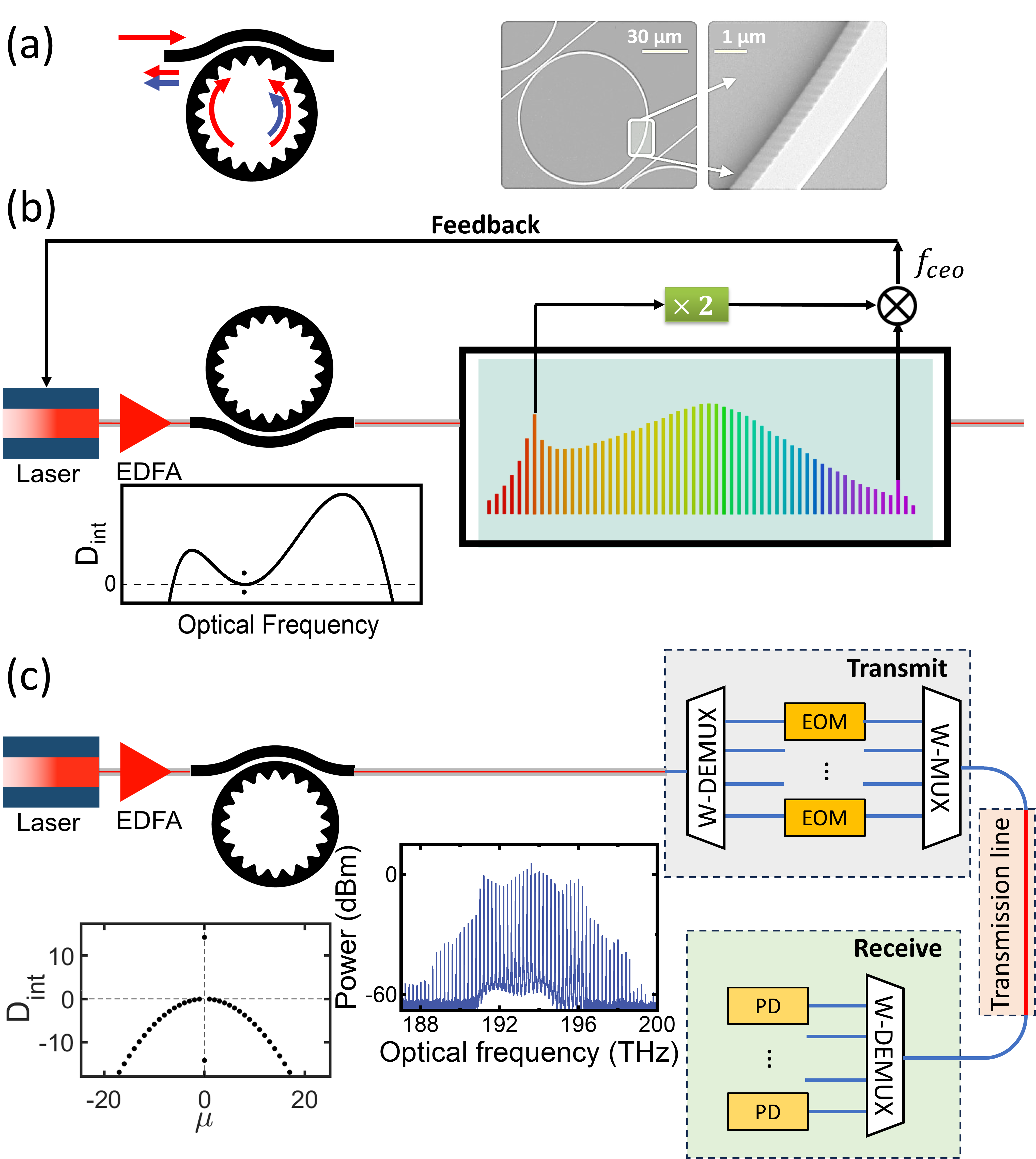}
\caption{\textbf{An introduction and applications of PhCRs.} (a) A diagram of PhCR and its SEM photos. The red arrows depict the pump wave flow and the blue arrows show the wave of the generated soliton (usually in the CCW direction). (b) A system diagram of an octave-spanning microcomb spontaneously generated in a PhCR and further used for f-2f self-referencing. The inset plot shows the anomalous GVD and the split pump mode. (c) A system diagram of optical data transmission network utilizing microcomb. The PhCR has a normal GVD and output a high power soliton, with its $D_\text{int}$ and the soliton spectrum plotted in the two insets. The output soliton is sent to a transmit module (the gray box) where comb lines are separated by a W-DEMUX, encoded by EOMs and recombined by a W-MUX. Then, the signal is transmitted to the receive module (the green box) where the comb lines are separated by another W-DEMUX and received by PDs.}
\label{fig1}
\end{figure}

A photonic-crystal (PhC) is a dielectric structure characterized by a refractive index that varies periodically at subwavelength scales. In microresonators, it can be implemented in the form of a sinusoidal (or other-shaped) modulation of the RW, resulting in a spatially varying effective refractive index. A PhC with an even repetition $m_\text{PhC}$ will induce scattering and couple the clockwise (CW) and counterclockwise (CCW) propagating waves inside the resonator with mode number $m = m_\text{PhC}/2$ and split it into two resonance modes \cite{liu2025implementing}. In the most convenient applications, we implement PhC on the pump mode, providing the flexibility of phase matching to any other modes. Figure \ref{fig1}(a) shows the diagram and SEM images of a PhCR. When the pump laser enters the PhCR, it will propagate in both CW and CCW directions, depicted by the red arrows. As we increase the pump power, both numerical simulations and experiments show that, in most cases, the pump mode intensity in the CCW direction reaches the nonlinear threshold first, exciting the primary comb or solitons, labeled by the blue arrows. As discussed in the introduction, PhCRs enable the stable soliton generation in both anomalous and normal GVD regimes, providing great application prospects in various photonics systems. Figure \ref{fig1}(b) shows a system diagram of a photonics circuit generating phase-stabilized frequency comb. The pump laser is amplified by an erbium-doped fiber amplifier (EDFA) and injected into an anomalous GVD PhCR with a split pump mode. The well-designed PhCR enables the spontaneous generation of a broadband soliton with two DWs spanning an octave. By frequency doubling the LWDW and heterodyning it with the SWDW, we can measure the $f_\text{ceo}$ and stabilize it through feedback control. More applications can be found in PhCRs with normal GVD. The normal GVD solitons are characterized by high power and a flat-top spectrum of which comb lines become good candidates for channels in data transmission. This scheme was demonstrated in experiments in recent years \cite{yang2022multi}. Figure \ref{fig1}(c) shows a diagram of its simplified version. The soliton excited by a normal GVD PhCR with pump mode split is input into the transmit module (grey background). In the module, different comb lines are extracted by a wavelength demultiplexer (W-DEMUX), encoded by individual electro-optics modulators (EOM), and combined together by a wavelength multiplexer (W-MUX). After transmitting to the receive module through optical fibers, individual comb lines are extracted by another W-DEMUX and received by individual photodetectors (PDs). Recently, a 2.4-THz bandwidth optical coherent receiver has been realized utilizing PhCR microcombs, highlighting the prospects of PhCRs in optical data communication \cite{deakin20242}.   

\subsection{Characterization of PhCR Mode Split}
\begin{figure}[ht]
\centering
\includegraphics[width=\linewidth]{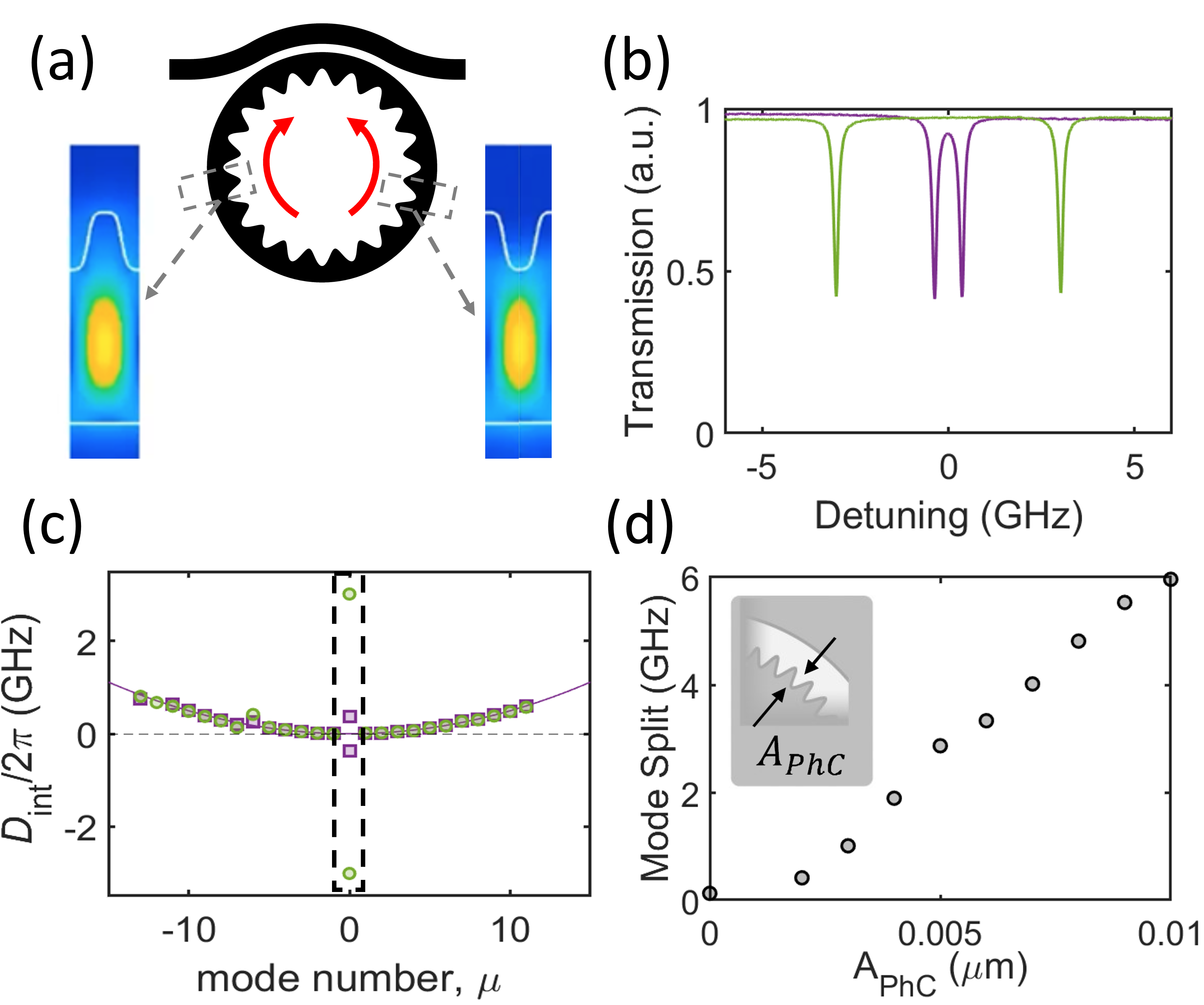}
\caption{\textbf{Characterizing PhCRs in experiments.}(a) The simulated mode profiles of the split resonance modes due to the sidewall modulation. (b) The measured transmission traces around the split modes from two PhCRs with $A_\text{PhC}$ = 2 nm (purple) and 10 nm (green). (c) The measured $D_\text{int}$ of two PhCRs with identical ring dimensions but different $A_\text{PhC}$. The purple squares correspond to the one with $A_\text{PhC}$ = 2 nm. The green circles correspond to the one with $A_\text{PhC}$ = 10 nm. The purple curve is the simulated $D_\text{int}$. The dashed box highlights the split pump modes. (d) The measured mode splits on PhCRs with different $A_\text{PhC}$. The inset shows the definition of $A_\text{PhC}$.}
\label{fig2}
\end{figure}

While PhCs result in distinctive dynamics of microresonators, it is essential to know how to characterize and manipulate them in design and experiment. In a PhCR, the CW and CCW propagating waves will be coupled together, resulting in two possible standing-wave resonance modes of which profiles are shown in Fig. \ref{fig2}(a). The difference between these two resonance frequencies is called the mode split $\Gamma$. From a theoretical point of view, $\Gamma$ also characterizes the strength of coupling between CW and CCW waves. We measure $\Gamma$ by sweeping the pump detuning in the experiment and recording the transmission trace. Figure \ref{fig2}(b) shows the transmission traces of two PhCRs with the same geometric parameters (RW = 1.55 $\mu$m) but two different $A_\text{PhC}$ values: 2 nm (purple) and 10 nm (green). For each device, the previous resonance dip is split into two dips and the distance between the two dips is $\Gamma$. The two pairs of resonance dips have different $\Gamma$s but similar depths, indicating a similar coupling coefficient $K$. Figure \ref{fig2}(c) further plots their corresponding $D_\text{int}$. The curve shows the simulated value. The purple squares are the measured $D_\text{int}$ of the PhCR with $A_\text{PhC}$ = 2 nm and the green circles are $D_\text{int}$ of the PhCR with $A_\text{PhC}$ = 10 nm. The purple squares and the green circles overlap with the curve except for the pump mode. We further investigate the relation between $\Gamma$ and $A_\text{PhC}$ and plot the measured $\Gamma$ in Fig. \ref{fig2}(d). By scanning the $A_\text{PhC}$ from 2 nm to 10 nm, we can tune $\Gamma$ from 0.7 GHz to 6.0 GHz. Previous research shows that a large $\Gamma$ can be easily achieved with a larger $A_\text{PhC}$ without impairing the quality factor \cite{liu2025implementing}. Figure \ref{fig2} (b), (c) and (d) suggest that by adjusting $A_\text{PhC}$, we can control $\Gamma$ in a convenient way independent of the overall dispersion profile and the coupling parameters. 

\section{Soliton Generation in PhCR}
The most extraordinary capability of PhCR is the stable access to solitons. Previous research has suggested that the soliton stability is affected by the relation between the branches of the continuous-wave solutions of a nonlinear system and the threshold of the primary comb \cite{liu2025exploring}. The fields inside PhCRs are governed by the coupled LLE \cite{skryabin2020hierarchy}:
 \begin{equation}
 \begin{split}
 \frac{\partial\psi_{\text{t}}}{\partial\tau} = &-(1+i\alpha)\psi_{\text{t}} + \frac{iD_2}{\kappa}\frac{\partial^2\psi_{\text{t}}}{\partial\theta^2} + i(|\psi_{\text{t}}|^2 + 2|\psi_{\text{r}}|^2_\text{avg})\psi_{\text{t}} \\&- i\frac{2\pi\Gamma}{\kappa}(\psi_{\text{r}})_\text{avg} + F\\
 \frac{\partial\psi_{\text{r}}}{\partial\tau} = &-(1+i\alpha)\psi_{\text{r}} + \frac{iD_2}{\kappa}\frac{\partial^2\psi_{\text{r}}}{\partial\theta^2} + i(|\psi_{\text{r}}|^2+2|\psi_{\text{t}}|^2_\text{avg})\psi_{\text{r}} \\&- i\frac{2\pi\Gamma}{\kappa}(\psi_{\text{t}})_\text{avg}.
 \end{split}
 \label{LLE}
 \end{equation}
Here, the subscripts t and r represent the CW and CCW propagating waves. $\theta$ is the angular coordinate. $\kappa$ is the resonator loss rate, and $\alpha$ is the pump detuning normalized to the resonator halfwidth $\kappa/(4\pi)$. $\tau$ is the time normalized to $2/\kappa$. $D_2$ is the second-order dispersion coefficient, and we assume $D_\text{int}=D_2\mu^2/(4\pi)$. $F$ is the normalized pump field. The subscript avg means taking the average over $\theta$. Compared with ordinary LLE, the extra photonic-crystal coupling terms $- 2\pi i\Gamma/\kappa(\psi_{\text{t/r}})_\text{avg}$ result in new dynamics of PhCRs and provide a different set of continuous-wave solutions. On the other hand, the threshold of primary combs with mode number $\mu$ and $-\mu$ in each direction has been well studied by previous research and calculated by the following equation \cite{liu2024threshold}:
 \begin{equation}
I_\text{t/r}^2 =1+\left(\alpha-2I+\frac{4\pi}{\kappa}D_\text{int}(\mu)\right)^2, 
 \label{threshold}
 \end{equation}
where $I_\text{t/r}$ is the intensity in each direction and $I$ is the total intensity of the two directions. Due to the opposite signs of $D_\text{int}$, the solitons in anomalous and normal GVD resonators have different waveforms. Therefore, we will analyze these two cases separately below. 
\subsection{Anomalous GVD case}
\begin{figure}[ht]
\centering
\includegraphics[width=\linewidth]{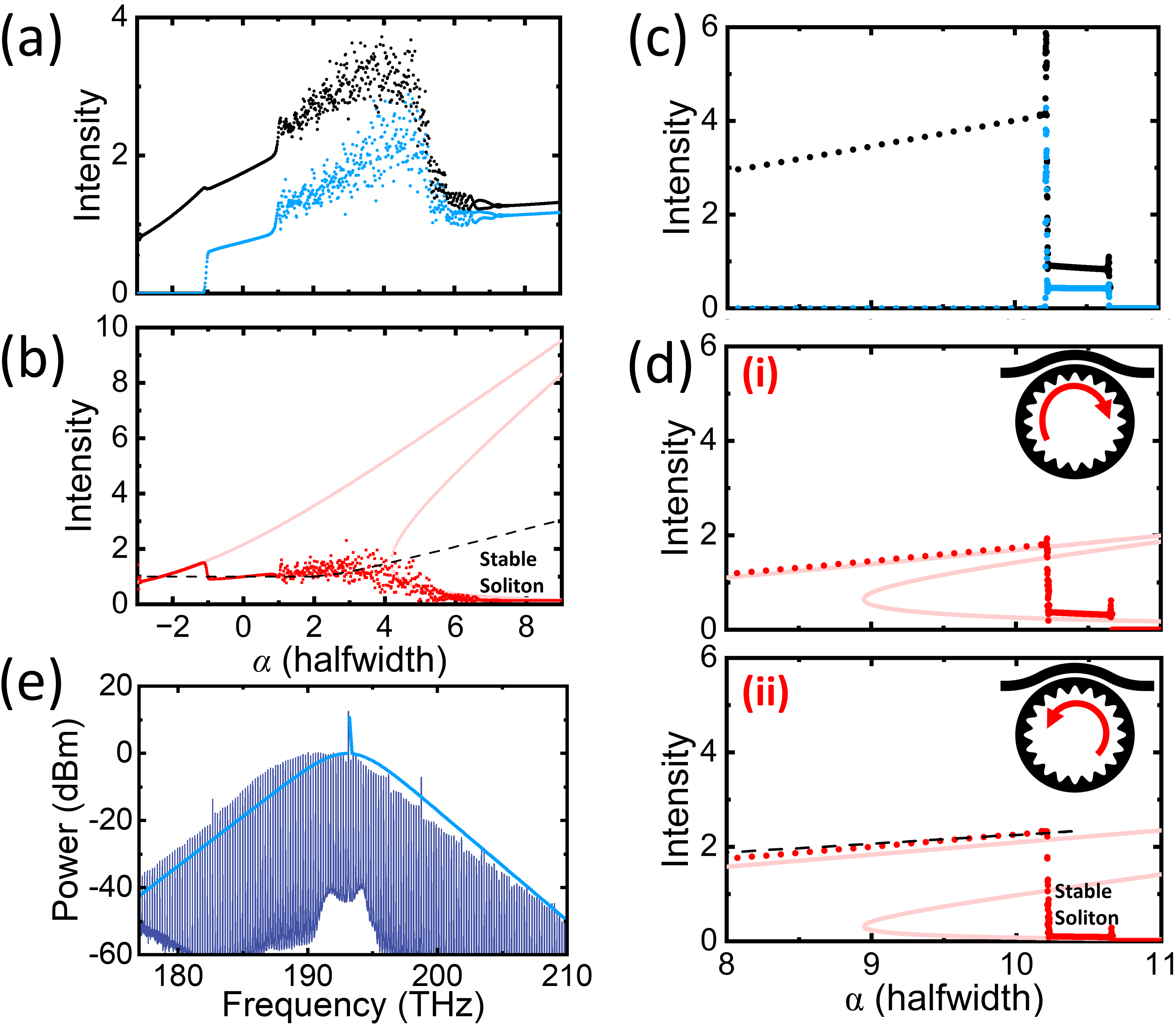}
\caption{\textbf{Dynamics of soliton formation in anomalous GVD microresonators.} (a) The total intracavity intensities (black trace) and the comb field intracavity intensity (blue trace) during the detuning sweep inside an ordinary microresonator with $F$ = 3.5. (b) The stability analysis of an ordinary microresonator during the detuning sweep. The pink curves are branches of continuous-wave solutions. The black dashed line plots the upper boundary of the region where the continuous wave is stable. The red trace plots the intracavity of the pump mode in the simulation. Same simulation condition as (a). (c) The total intracavity intensities (black trace) and the comb field intracavity intensity (blue trace, CCW direction) during the detuning sweep inside a PhCR with $\Gamma$ = 10.34 halfwidth and $F$ = 3.5. (d) The stability analysis of a PhCR in (i) CW and (ii) CCW directions. The pink curves plot the continuous-wave solutions. The red traces are the pump intensity during the simulation. The black dashed line is the upper boundary of the stable region of continuous wave. Same simulation condition as (c). (e) A comparison between the soliton spectra in the simulation (light blue trace) and in the experiment (dark black trace). The soliton propagates in the CCW direction.}
\label{fig3}
\end{figure}
The solitons in anomalous GVD resonators have a signature of narrow pulse shape that can be approximated by the sech$^2$ function \cite{zang2024foundry}. The sharp peak of this pulse is a result of the balance between the Kerr shift due to the high intensity at the peak and the second order dispersion due to the large curvature at the peak. However, the chaotic regime that precedes soliton existence poses a significant challenge to experimental generation of solitons without additional techniques. Here, we perform numerical simulations, based on the ordinary LLE, and investigate the intracavity intensity of the comb field and the pump field \cite{zang2024foundry}. Figure \ref{fig3}(a) plots the intensity of the comb field (blue trace) and the total intensity (including the pump field, black trace) over detuning $\alpha$. We observe a large fluctuation of the total intensity during the chaotic stage, which will result in thermal instability in experiments and prevent soliton formation. To explain the occurrence of the chaotic regime, we provide a necessary condition for a stable soliton. Considering that the intensity of the soliton pulse decays exponentially from the peak, the field at the remaining part of the resonator can be approximated by a continuous wave. On the other hand, since the pulse is sharp and exists locally, the pump intensity can also be approximated by this continuous intensity. Therefore, a necessary condition for stable soliton formation is that there exists a continuous-wave solution with intensity below the threshold of the primary comb. Otherwise, multiple primary combs will be excited and the system will go into chaos. Combining Eqn. \ref{threshold} with the condition $D_\text{int}\ge0$, we can derive the condition for the generation of the primary comb in anomalous GVD microresonators:
 \begin{equation}
 \begin{cases}
\alpha\le2I + \sqrt{I_\text{t/r}^2-1}\\
I_\text{t/r}\ge1
\end{cases}.
 \label{anomalousthre}
 \end{equation}
To verify our theory, in Fig. \ref{fig3}(b) we record the pump mode intensity in this simulation (red trace) and compare it with the analytic solutions of the continuous wave (pink traces). We also plot the boundary of the region where the threshold of primary comb can be reached in a black dashed line, based on Eqn. \ref{anomalousthre}. The light propagates in one direction in the ordinary microresonator so that $I=I_\text{t}$. For the region above this black dashed line, primary combs exist. When $\alpha\in[-2,4]$, there is only one continuous solution and it is above the threshold. Accordingly, a primary comb occurs first and then the system goes into chaos. When $\alpha>4$, two other branches of continuous-wave solutions occur. The simulated pump intensity gradually approaches the lowest branches, corresponding to the formation of a stable soliton.

From the analysis above, we know that in order to avoid the occurrence of a chaotic regime, the lowest branch of the continuous-wave solution needs to occur before the highest branch reaches the threshold of the primary comb \cite{liu2025exploring}. This condition cannot be satisfied in ordinary microresonators but can be achieved in PhCRs. To illustrate it, we numerically simulate the soliton generation process in PhCR based on Eqn. \ref{LLE}. Figure \ref{fig3}(c) plots the total intensity (black trace) and the comb intensity (blue trace). The comb was generated without a chaotic regime in the CCW propagating direction. We further plot the analytic solutions of continuous waves for CW and CCW directions (pink traces) in Fig. \ref{fig3}(d)(i) and (ii), and compare them with the recorded pump intensities (red traces) in the simulation. We also calculate the boundary of the region where the primary comb exists (black dashed line), based on the simulated $I_\text{t}$ and $I_\text{r}$. It can be seen from Fig. \ref{fig3}(d)(ii) that the pump intensity in the PhCR reaches threshold at a much larger detuning $\alpha=10$ compared with the case of the ordinary microresonator because the PhC red shifts the resonance frequency of the pump mode. The lower branch of the continuous-wave solution has already occurred before this detuning, so that the primary comb forms directly into a stable soliton, characterized by a jump in the pump intensity from the higher branch to the lower branch. We further verify our simulation with experiments and successfully generate a soliton in a PhCR device without extra techniques. Figure \ref{fig3}(e) compares the simulated soliton spectrum (light blue trace) and the measured spectrum (dark blue trace), which shows good agreement. The simulation takes $F = 3.5$ and the experimentally measured $\Gamma $ = 1.3 GHz (10.34 halfwidth).

\subsection{Normal GVD case}
\begin{figure}[!ht]
\centering
\includegraphics[width=\linewidth]{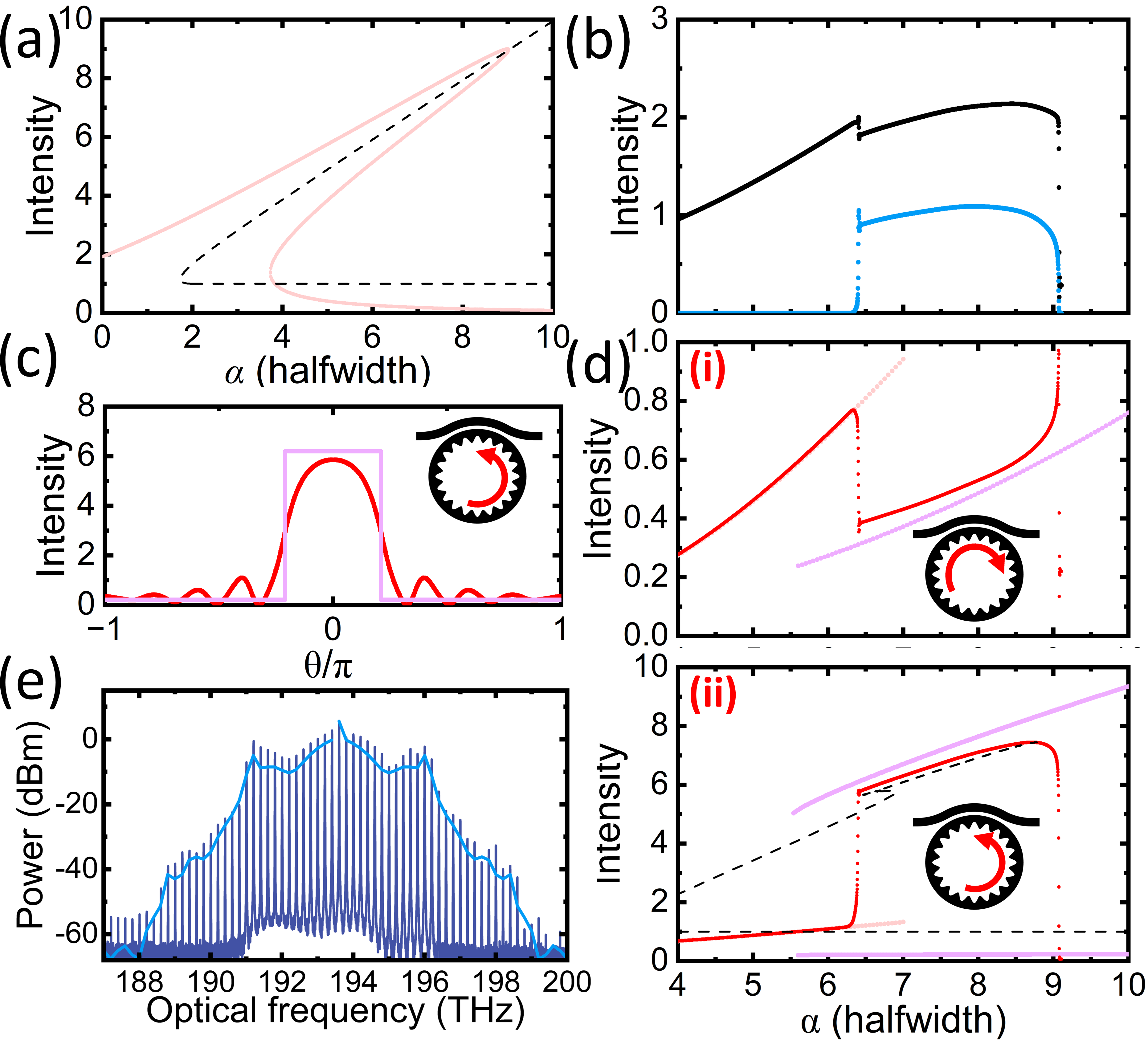}
\caption{\textbf{Dynamics of soliton formation in normal GVD microresonators.} (a) The pink curves plot the continuous-wave solution inside an ordinary microresonator with $F$ = 3. The area between the two black dashed lines are the instable region where the primary comb can form. (b) The total intracavity intensities (black trace) and the comb field intracavity intensity (blue trace, CCW direction) during the detuning sweep inside a PhCR with $\Gamma$ = 9.3 halfwidth and $F$ = 3. (c) The red curve are the soliton waveform in a normal GVD PhCR in the simulation at $\alpha$ = 6.5. The magenta line plots the square wave used to approximate the soliton waveform, with the high and low domains to be the highest and lowest continuous-wave solutions. Same simulation condition as (b). (d) The intracavity intensity analysis of a PhCR in (i) CW and (ii) CCW directions. The pink curves plot the intensity solutions when both CW and CCW directions have continuous waves. The magenta curve in (i) plots the intensity in the CW direction according to the Maxwell point, and the magenta curves in (ii) plot the highest and lowest branches of continuous-wave solutions in the CCW directions at the Maxwell point. The red traces plot the peak intensity of the field in the coordinate space in both directions. The area between two black dashed line in (ii) is the region where continuous-wave solutions are not stable. The simulation condition is same as (b). (e) A comparison between the soliton spectra in the simulation (light blue trace) and in the experiment (dark blue trace). The soliton propagates in the CCW direction.}
\label{fig4}
\end{figure}

The condition for the generation of primary comb in a normal GVD microresonator is quite different from the anomalous case due to the negative $D_\text{int}$. Based on Eqn. \ref{threshold}, it can be reduced to:
 \begin{equation}
 \begin{cases}
\alpha\ge2I - \sqrt{I_\text{t/r}^2-1}\\
I_\text{t/r}\ge1
\end{cases}.
 \label{normalthre}
 \end{equation}
It is impossible to excite the primary comb in an ordinary microresonator with normal GVD by gradually sweeping the detuning of the continuous wave pump from low to high. Figure \ref{fig4}(a) shows the reason. The pink trace is the continuous wave solution of an ordinary microresonator with $F = 3$. The region between the two black dashed lines is the existence range of the primary comb, depicted by Eqn. \ref{normalthre}. As we sweep the detuning from low to high in experiments, the system will evolve along the highest branch of the pink trace and drop to the lowest branch without any chance to fall between the two black dashed lines. From a physical point of view, the pump intensity increases as the detuning increases, which red-shifts the resonance frequency of comb modes further due to the Kerr effect, preventing OPO from happening. However, these continuous-wave solutions are modified in a resonator with coupled bidirectionally propagating waves, enabling the excitation of the primary comb and also the soliton. Some recent research has studied and demonstrated experimentally such soliton generation  by utilizing a self-injection locking microresonator system which also includes the bidirectionally propagating field \cite{PhysRevA.106.053508}. Here, we will apply some of its analysis to the PhCRs. 

We numerically simulate the generating process of the primary comb and soliton inside a PhCR with $F=3$ and $\Gamma = 0.9$ GHz (9.3 halfwidth). Figure \ref{fig4}(b) plots the simulated intensity of comb (light blue trace) that propagates in CCW direction and the total intensity inside the resonator (black trace). Different from the sharp peak in the anomalous GVD cases, the soliton in normal GVD resonators has a much flatter top because the dispersion and Kerr shift cannot balance each other. Its waveform is usually approximated by a square wave with the high-intensity domain and low-intensity domains being the highest and the lowest branches of the continuous-wave solutions and connected by switching waves. Previous research indicates that the bidirectionally propagating waves are self-regulated around the Maxwell point characterized by $f\approx4\alpha/\pi^2$, where $f$ is the effective pump field in the direction where the soliton exists \cite{PhysRevA.106.053508}. In our case, the soliton exists in the CCW direction. From the second equality in Eqn. \ref{LLE} we know that the photonic-crystal interaction term $- 2\pi i\Gamma(\psi_{\text{t}})_\text{avg}/\kappa$ acts as $f$. With this insight, when the soliton has already formed in the CCW direction, the intensity of the continuous wave in the CW direction can be approximated by:
\begin{equation}
I_\text{t}=\left(\frac{\kappa}{2\pi\Gamma}\right)^2\frac{16\alpha^2}{\pi^4}.
\label{It}
\end{equation}
We compare the intensity calculated by this analytical model with our simulation to check the validity. The red trace in Fig. \ref{fig4}(c) plots the soliton shape in the simulation at $\alpha = 6.5$, and the magenta trace plots a square wave of which the high and the low intensity domains are the highest and lowest branches of the continuous-wave solution of the second equality of Eqn. \ref{LLE} and Eqn. \ref{It}. The higher analytical solution is quite close to the top of the soliton, suggesting that the square wave is a good approximation of the soliton waveform in a normal GVD resonator. We further study the complete evolution of the intra-cavity field from continuous waves in both directions, to the primary comb in the CCW direction and finally to the soliton in the CCW direction. The red traces in Fig. \ref{fig4}(d)(i) and (ii) record the peak intensity of the field in coordinate space in CW and CCW directions, respectively. The pink curves are the analytical solutions of Eqn. \ref{LLE}, assuming the field in both directions are continuous waves. The magenta curve in Fig. \ref{fig4}(d)(i) is the approximated $I_\text{t}$ according to Eqn. \ref{It}, and the two magenta curves in Fig. \ref{fig4}(d)(ii) are the highest and lowest intensity domains of the square wave used as the approximate waveform of the soliton. The region between the two black dashed traces in Fig. \ref{fig4}(d)(ii) is the region where the primary comb can exist. The boundaries are calculated based on Eqn. \ref{normalthre} and the simulated $I_\text{t}$ and $I_\text{r}$. Now, we are going to explain the evolution of the fields in detail. Focusing on Fig. \ref{fig4}(b)(ii), when $\alpha<5.5$, the pink trace is below the region between the two black dashed lines, meaning that the threshold of the primary comb has not been reached. Accordingly, the fields in both directions are continuous waves, verified by the overlapping between the red traces and the pink traces. When $\alpha\in[5.5,8.7]$, the pink curve goes between the two dashed lines. The primary comb is excited in the CCW direction, verified by the deviation of the red trace from the pink curve. As the red trace goes above the upper black dashed line, a soliton is formed in the CCW direction. In this regime, the red trace in Fig. \ref{fig4}(d)(i) is quite close to the magenta curve characterized by the Maxwell point, and the red trace in Fig. \ref{fig4}(d)(ii) approaches the magenta curve, which plots the higher intensity domain of the square wave used for approximation, verifying the validity of our analytical model. The fact that the red trace is above the higher black dashed line also indicates that the higher intensity domain of this soliton is stable until the red trace intersects with the black dashed line at $\alpha=8.7$, corresponding to the collapse of the soliton. We further verify our theory with the experiment. The dark blue trace in Fig. \ref{fig4} plots the measured spectrum of a soliton output in the CCW direction from a PhCR with $\Gamma=0.9$ GHz (9.3 halfwidth). The light blue trace is the simulated spectrum using the same parameters, showing a good agreement with the experimental result. 

\section{Nanocomposite Microresonator Network for Pump harmonic microcomb}
Self-referenced microcombs provide a convenient on-chip frequency standard for optical metrology. As discussed in the introduction, the traditional self-referencing using octave spanning comb suffers from the limited power of the LWDW during the frequency doubling. In this section, we will explore the possibility of using a pump-harmonic microcomb as a next generation phase-stabilized soliton frequency comb, which uses the pump wave and the SWDW at the doubled frequency of the pump wave to do the self-referencing. To maximize the power of SWDW and stabilize the soliton generation, we will introduce below a promising scheme of a nanocomposite microresonator network.
\subsection{GVD Advantage of Nanocomposite Microresontor}
\begin{figure}[ht]
\centering
\includegraphics[width=\linewidth]{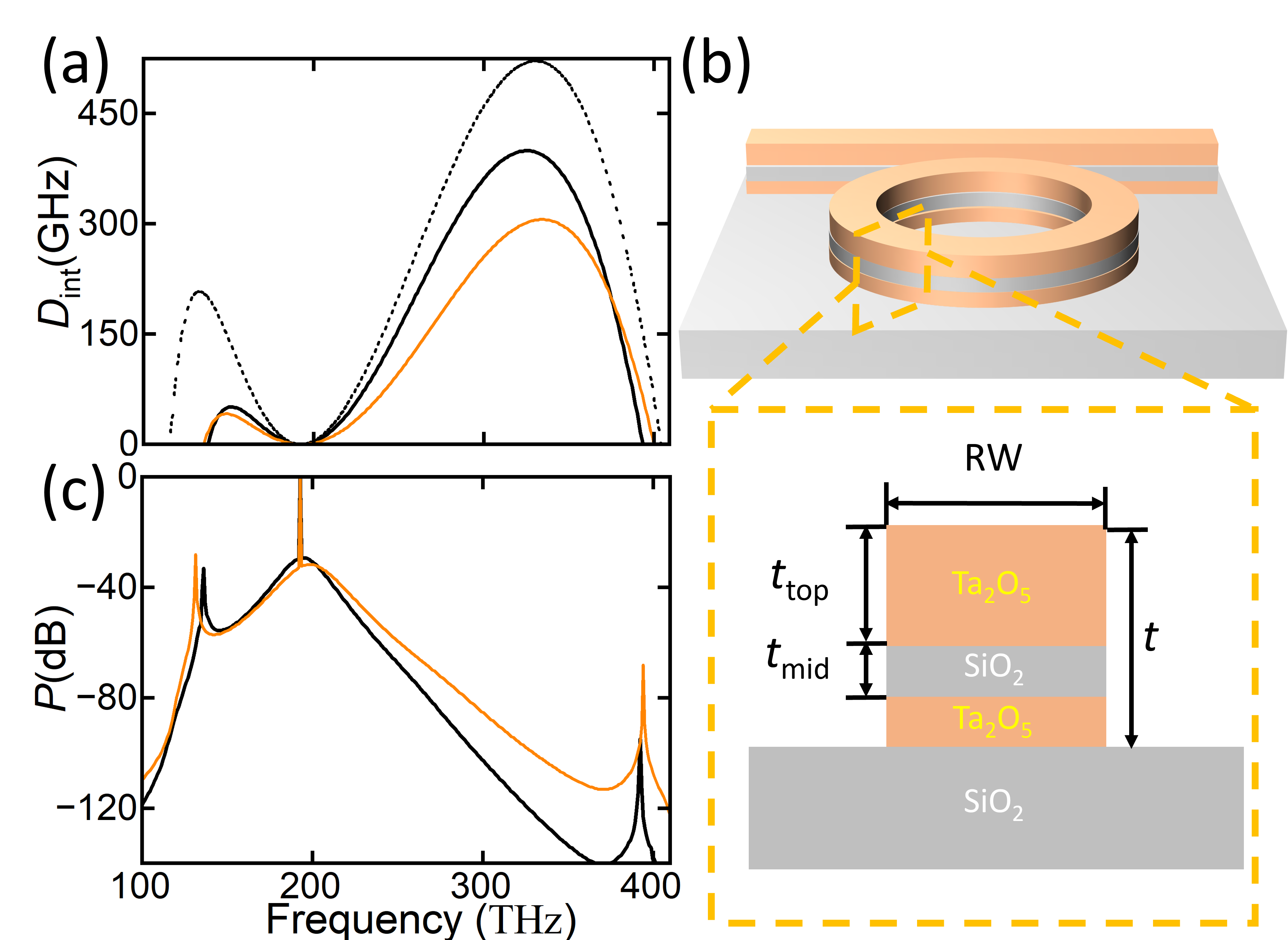}
\caption{\textbf{Comparison of GVD in single layer and nanocomposite microresonators.} (a) The solid black curve plots the simulated $D_\text{int}$ of a resonator with ($t$, RW) = (0.55, 1.307) $\mu$m. The dashed black curve plots the simulated $D_\text{int}$ of a resonator with ($t$, RW) = (0.6, 1.363) $\mu$m. The orange curve plots the simulated $D_\text{int}$ of a nanocomposite resonator with ($t$, $t_\text{top}$, $t_\text{mid}$, RW) = (1.15, 0.6, 0.25, 1.082) $\mu$m. (b) A diagram of the nanocomposite microresonator. The upper and lower layers are made of Ta$_2$O$_5$ and the middle layer is made of SiO$_2$. (c) The simulated soliton spectra output from the single layer (black) and nanocomposite (orange) resonators, based on the corresponding $D_\text{int}$ in (a). $F$ = 10 in the simulation.}
\label{fig5}
\end{figure}
The spectrum of a soliton microcomb is mostly affected by the $D_\text{int}$. For the DW solitons in an anomalous GVD resonator, the three most critical indexes of $D_\text{int}$ are the frequency of SWDW ($\nu_\text{S}$) and the frequency of LWDW ($\nu_\text{L}$), which correspond to $D_\text{int}=0$, and the second-order dispersion parameter $D_2$. In the case of pump-harmonic microcombs, $\nu_\text{S}$ should be targeted at twice the pump frequency ($\nu_\text{0}$). Previous research demonstrated that a smaller $D_2$ fundamentally leads to the higher SWDW power, while the smaller $\nu_\text{L}$ mitigates the nonlinear self-interaction that causes the instability and collapse of the soliton \cite{zang2024foundry,skryabin2017self}. For an ordinary waveguide, the only two geometric parameters are $t$ and RW. As a result, $D_2$ and $\nu_L$ cannot change independently  with the constraint $\nu_\text{S}=2\nu_0$. Figure \ref{fig5}(a) shows the tradeoff between $D_2$ and $\nu_L$ for ordinary waveguides. The solid and dashed black curves are the simulated $D_\text{int}$ using COMSOL of two tantala waveguides on silica bases with ($t$, RW) = (0.55, 1.307) $\mu$m and (0.6, 1.363) $\mu$m. If we fix $\nu_\text{S}=2\nu_0$ and reduce $\nu_\text{L}$ by increasing $t$, we cannot avoid increasing $D_2$. However, for our nanocomposite waveguide design that sandwiches a silica layer with two tantala layers (Fig. \ref{fig5}(b)), we achieve a higher level of control of the profile of the thickness, namely the total thickness $t$, the middle silica layer thickness $t_\text{mid}$ and the top tantala layer thickness $t_\text{top}$ enable us to reduce $\nu_\text{L}$ and $D_2$, simultaneously. The orange curve in Fig. \ref{fig5}(a) is the simulated $D_\text{int}$ of a nanocomposite structure with ($t$, $t_\text{top}$, $t_\text{mid}$, RW) = (1.15, 0.6, 0.25, 1.082) $\mu$m. Compared with the solid black curve, both $\nu_\text{L}$ and $D_2$ are improved (reduced). We further numerically calculate the soliton spectrum solutions of the Lugiato-Lefever equation (LLE) (see Fig. \ref{fig5}(c)) using the solid black and orange $D_\text{int}$ curves in Fig. \ref{fig5}(a). The soliton spectrum in the nanocomposite waveguide (orange) has overall higher power and >20 dB increase at the SWDW power. 

\begin{figure}[ht]
\centering
\includegraphics[width=\linewidth]{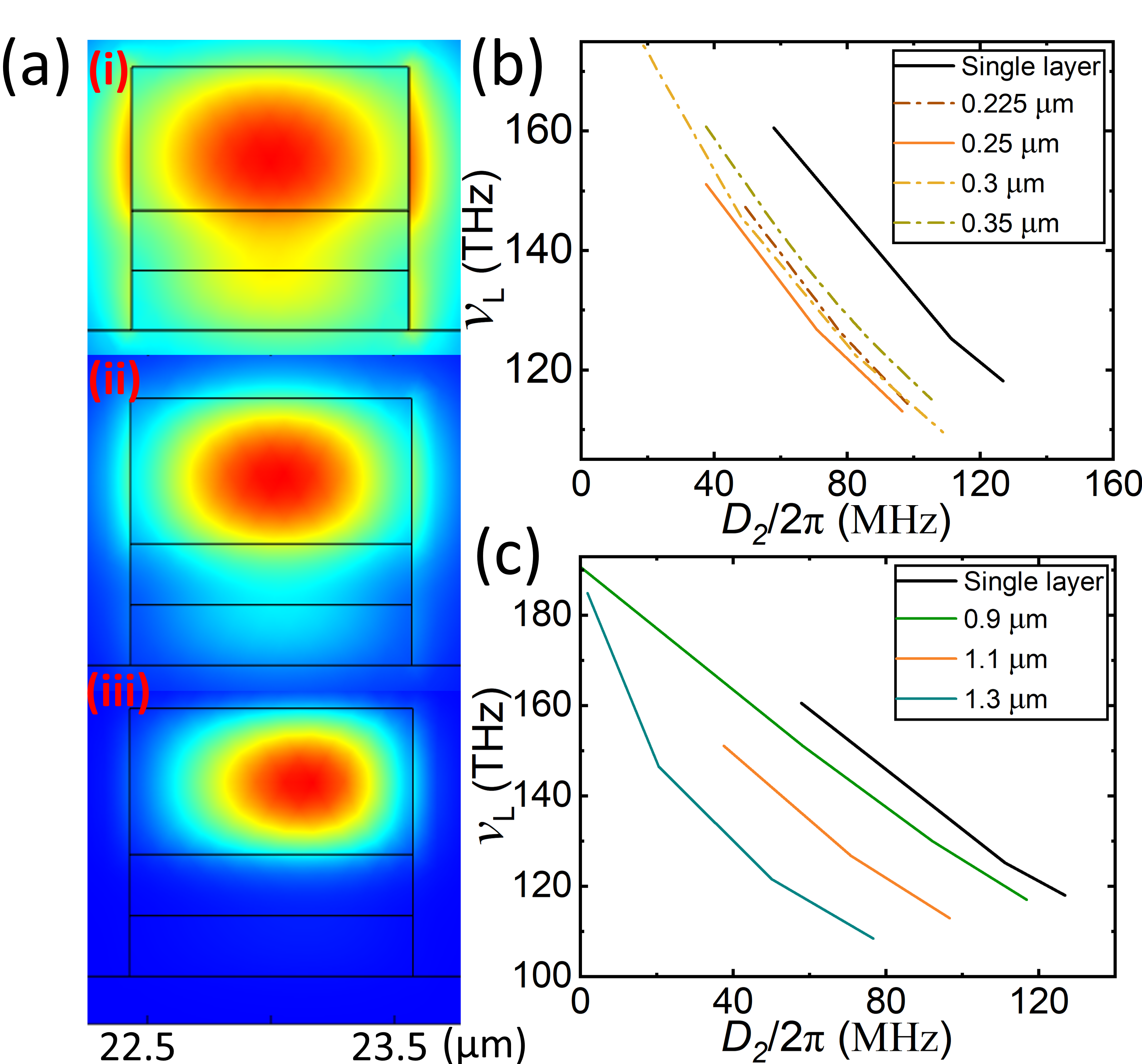}
\caption{Optimization of layer thicknesses of nanocomposite waveguides. (a) The mode profiles in the same nanocomposite waveguide with different frequencies: (i) 127 THz, (ii) 193 THz, (iii) 386 THz. (b) The $D_2$ - $\nu_\text{L}$ dependence in single layer waveguide and nanocomposite waveguides with $t$ = 1.1 $\mu$m and different values of $t_\text{mid}$. $\nu_\text{S}$ is fixed at 386 THz. (c) The optimal $D_2$ - $\nu_\text{L}$ boundary with single layer waveguide and nanocomposite waveguides with different values of $t$. $\nu_\text{S}$ is fixed at 386 THz.}
\label{fig6}
\end{figure}

Now, we explain the physical mechanism of the fundamental advantage in a nanocomposite resonator, specifically that pump-harmonic combs can be realized at lower $D_2$. We plot the frequency-dependent mode profiles in a nanocomposite waveguide at three different frequencies ($\nu_\text{L}$, $\nu_\text{0}$ and $\nu_\text{S}$) in Fig. \ref{fig6}(a). At $\nu_\text{L}$ (panel (i)), the mode can almost permeate all three layers. At $\nu_0$ (panel (ii)), the mode partially permeates into the bottom layer. At $\nu_\text{S}$ (panel (iii)), the mode is mostly confined in the top layer. Therefore, the middle layer results in a frequency-varying effective waveguide dimension. The mechanism does not depend on the ring structure and can also be applied to the dispersion optimization of straight waveguides. To quantitatively investigate its effect, we fix $t$ = 1.1 $\mu$m, $\nu_\text{S}$ = 386 THz and for each $t_\text{mid}$ we sweep the $t_\text{top}$ to obtain the $D_2$ - $\nu_\text{L}$ dependence (see Fig. \ref{fig6}(b)). The result shows that there exists an optimal $t_\text{mid}$ value (250 nm for $t$ = 1.1 $\mu$m) corresponding to the left lower boundary in the $D_2$ - $\nu_\text{L}$ plot. Figure \ref{fig6}(c) plots the shift of this boundary with different $t$ values. Among the range we are interested in, a larger $t$ potentially provides better dispersion configurations but also causes challenges in the etching process during fabrication.

\subsection{Soliton generation in microresonator network}
\begin{figure}[ht]
\centering
\includegraphics[width=\linewidth]{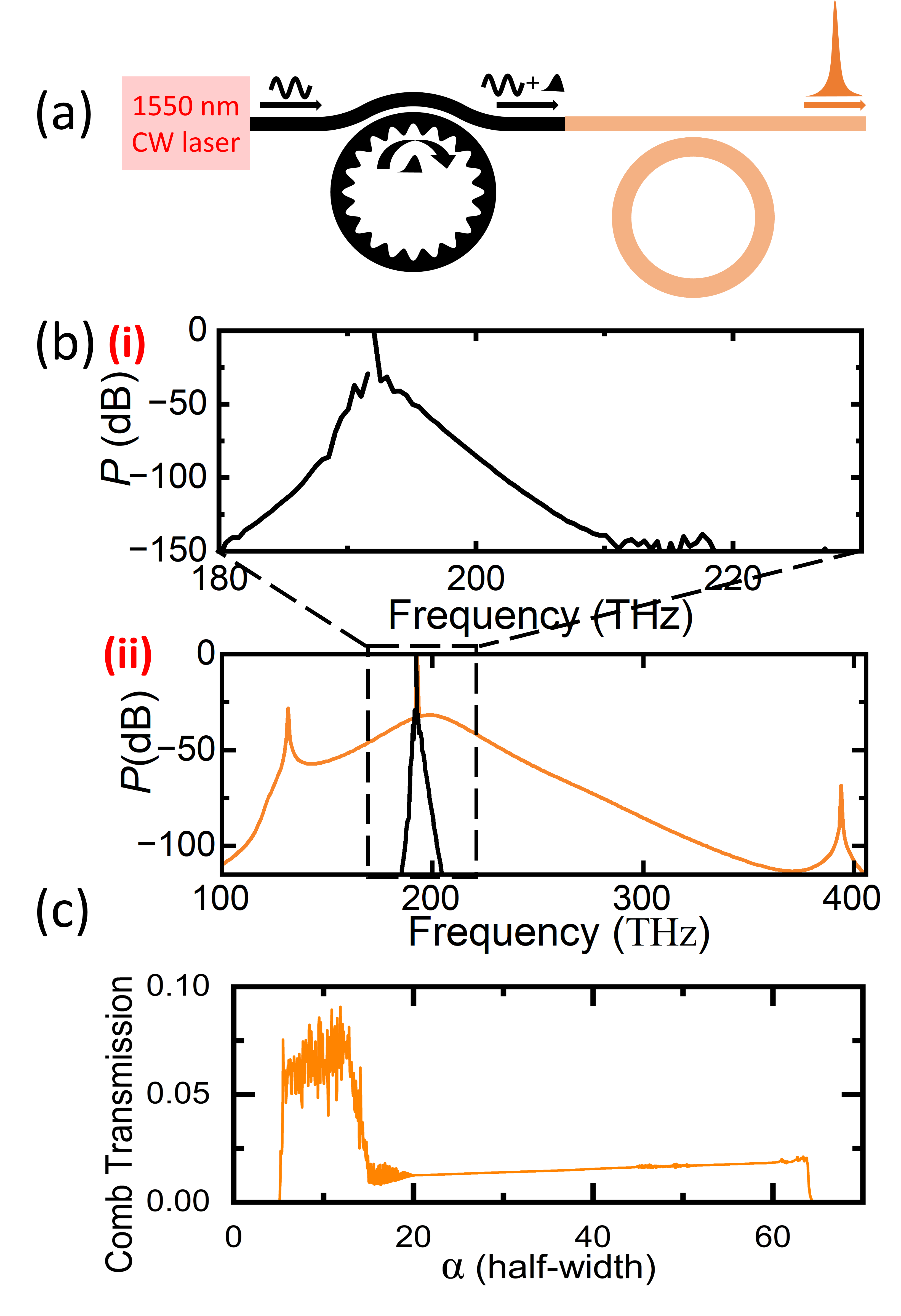}
\caption{\textbf{The nanocomposite microresonator network.} (a) A system diagram of the two-microresonator network. The black part is a nanocomposite normal GVD PhCR with mode $\mu$ = -1 split that converts the continuous laser into a pulse pump. The orange part is the nanocomposite anomalous GVD microresonator used to generate pump harmonic microcomb. (b) The spectrum of waves output from the PhCR (black curve) and the nanocomposite microresonator (orange curve). (c) The simulated comb transmission of the network over the detuning $\alpha$ (normalized by the halfwidth of the second ring).}
\label{fig7}
\end{figure}
The nanocomposite structure waveguide provides favorable GVD for the pump-harmonic microcomb, but how to excite the soliton in the microresonator remains a question. While this can be achieved by adding a PhC to split the pump mode, recent experiments show that the sidewall modulation will cause much scattering loss around the frequency much higher than the pump \cite{pimbi2025spectralresponsegratinginducedloss}. For high frequencies, the grating introduces scattering loss rather than coherent reflection, leading to significantly reduced $Q$ by several orders of magnitude compared to the pump, and thus reduces the power of SWDW. In the past few years, pulse pumping has been investigated as an alternative way to continuous-wave pumping to excite solitons in microresonators \cite{Anderson_2021}. When the difference between the FSR of the microresonator and the repetition rate of the pulse pump is small enough, the soliton in the microresonator will lock to the driving pulse \cite{Obrzud_2017}. 

With this insight, we propose an integrated microresonator network for the pump harmonic microcomb generation with continuous-wave pump. In this scheme, the continuous wave will excite a narrowband pulse in the first microresonator, and then this pulse together with the transmitted continuous-wave light will pump the second microresonator and excite the pump harmonic microcomb. Figure \ref{fig7}(a) shows a diagram of this network. The black part is a nanocomposite bandgap-detuned normal PhCR.
To be specific, the mode split is implemented on $\mu$ = -1 mode rather than the pump mode, which will yield a pulse in the CW direction \cite{Jin_2025}. The orange part is a nanocomposite anomalous microresonator for SWDW excitation. The geometry of the second microresonator is optimized to give the most favorable GVD for the highest SWDW power. The first microresonator shares the same layer thicknesses but the RW and RR are adjusted to yield the same FSR and pump resonance frequency as the second microresonator. To check the feasibility of this scheme, we numerically simulate the dynamics of this system. Figure \ref{fig7}(b)(i) shows the simulated output spectrum of the pulse and the transmitted continuous-wave light from the bandgap-detuned PhCR. Then, this spectrum is used as the pump input in the simulation of the second microresonator. The output spectrum of the second microresonator is shown in Fig. \ref{fig7}(b)(ii). Figure \ref{fig7}(c) plots the simulated comb transmission during the detuning sweep. The orange spectrum in Fig. \ref{fig7}(b)(ii) is very similar to the orange spectrum in Fig. \ref{fig6}(c), suggesting that once the soliton is generated, its spectrum is not determined by the pulse pump but determined by the GVD of the resonator. 

\section{Summary}
This article discusses prospects of generating soliton frequency combs in photonic-crystal and nanocomposite microresonator systems. The PhCRs enable stable soliton generation in both anomalous and normal GVD regimes, resulting in great applications in optical metrology and optical data transmission. To explain the soliton formation process, we perform a thorough theoretical analysis and our model matches with numerical simulations and experiments well, providing valuable guidance for the further parameter design and experiment. To overcome the drawback of traditional self-referencing octave-spanning combs, we propose the new concept of pump harmonic microcomb and the nanocomposite microresonators. The new waveguide structure releases extra degrees of freedom in dispersion engineering, yielding a much more favorable GVD for high power DWs excitation that cannot be achieved by any single-layer structure. Its advantage is explained by the distinctive frequency-dependent effective waveguide structure, and the middle layer plays a crucial role. More importantly, this revolutionary idea can also apply to the dispersion optimization of general waveguides. Finally, inspired by the previous research of pulse-driven microresonators, we propose a two-microresonator network, using a PhCR to convert the continuous wave pump laser into a pulse pump and excite broadband soliton in the second resonator with well-designed GVD. Our work highlights a path to achieve robust, compact next-generation soliton frequency combs that have great prospects in fundamental research and engineering.

\begin{backmatter}
\bmsection{Funding} 
AFOSR FA9550-20-1-0004 Project Number 19RT1019, NSF Quantum Leap Challenge Institute Award OMA – 2016244, DARPA PIPES and NaPSAC, and NIST. 
\bmsection{Acknowledgment} 
We thank Grant Brodnik and Lindell Williams for the technical review of the paper and Jennifer Black for discussions at the beginning of the work. Trade names provide information, not an endorsement.

\bmsection{Disclosures} The authors declare no conflicts of interest.

\bmsection{Data availability} Data underlying the results presented in this paper are not publicly available at this time but may be obtained from the authors upon reasonable request.

\end{backmatter}

\bibliography{sample}

\bibliographyfullrefs{sample}


\end{document}